\documentclass[12pt]{article}
\makeatletter
\def\@biblabel#1{#1.}
\makeatother
\usepackage{times}
\usepackage{bm}
\usepackage[dvips]{graphicx}
\newcounter{lastnote}

\title{Spin-Rotation Symmetry Breaking in the Superconducting State of  Cu$_x$Bi$_2$Se$_3$}

\author
{K. Matano, $^{1}$   M. Kriener,$^{2}$ K. Segawa,$^{2}$ Y. Ando$^{2,3}$ , and Guo-qing Zheng$^{1,4\ast}$\\
\\
\normalsize{$^{1}$Department of Physics, Okayama University, Okayama 700-8530, Japan}\\
\normalsize{$^{2}$Institute of Scientific and Industrial Research,}\\ 
\normalsize{ Osaka University, Osaka 567-0047, Japan}\\ %Ibaraki,
\normalsize{$^{3}$ Institute of Physics II, University of Cologne,}\\
\normalsize{ 50937 Cologne,Germany}\\ %Zuelpicher Str. 77,
\normalsize{$^{4}$Beijing National Laboratory for Condensed Matter Physics and Institute of Physics,}\\ 
\normalsize{Chinese Academy of Sciences, Beijing 100190, China}\\
\\
\normalsize{$^\ast$To whom correspondence should be addressed; E-mail:  zheng@psun.phys.okayama-u.ac.jp}
}

\date{}

\begin{document} 


\baselineskip24pt


\maketitle

% \begin{sciabstract}
% Abstract
% \end{sciabstract}
% 

% In setting up this template for *Science* papers, we've used both
% the \section* command and the \paragraph* command for topical
% divisions.  Which you use will of course depend on the type of paper
% you're writing.  Review Articles tend to have displayed headings, for
% which \section* is more appropriate; Research Articles, when they have
% formal topical divisions at all, tend to signal them with bold text
% that runs into the paragraph, for which \paragraph* is the right
% choice.  Either way, use the asterisk (*) modifier, as shown, to

%\section*{Introduction}
\textbf{
Spontaneous symmetry breaking is an important concept for understanding  physics ranging from the elementary particles to states of matter. % \cite{Nambu}.
For example, the superconducting state breaks global gauge symmetry, and  unconventional superconductors 
%such as cuprate high temperature superconductors, the spatial part  of the electron-pairs wave function 
can break additional symmetries. %  \cite{Sigrist}. 
%such as cuprate high temperature superconductors, not only the gauge-symmetry is broken but the spatial part  of the electron-pairs wave function has lower symmetry than that of underlying lattice .
In particular,  % with parallel spin pairs, 
 spin rotational symmetry is expected to   be broken in spin-triplet superconductors.  
 However, experimental evidence  for such %spin-rotation 
symmetry breaking has not been conclusively obtained
%has never been obtained 
so far in any candidate compounds. %,  which hinders understanding of  this novel class of superconductivity. % spin-triplet superconductivity \cite{UPtReview,MaenoReview,MaenoReview2}.  
Here, by $^{77}$Se nuclear magnetic resonance measurements, 
we show that  spin rotation symmetry is spontaneously broken in the hexagonal plane  of the electron-doped topological insulator Cu$_{0.3}$Bi$_2$Se$_3$ 
below the superconducting transition temperature $T_c$=3.4 K. %Furthermore,
Our results not only establish spin-triplet superconductivity in this compound, but may also serve to lay a foundation for the research of topological superconductivity.
 % provide the first evidence  for the long-sought symmetry breaking, % in a superconductor, 
%but 
 %also serve to lay a foundation toward establishing a new research frontier.
}
% class of superconductivity
%may also serve to lay a foundation toward  establishing a new class of superconductivity. %, topological superconductivity.  % indicate that Cu$_x$Bi$_2$Se$_3$  qualifies as a topological superconductor  
%that belongs to a   new class of materials %\cite{ZhangSC} 
%whose surface states have potential applications in quantum computing. %\cite{Kitaev}.

\newpage
In most superconductors including  copper-oxide and iron-pnictide high temperature superconductors, the electron pairs (Cooper pairs) are in the spin singlet state with total spin $S$=0 \cite{Tsuei,Matano}.  
Spin-triplet superconductivity or fluidity  with $S$=1 possesses internal structure, % (freedoms), % associated with the total spin, 
which gives rise to rich  physics such as additional symmetry breaking,  order parameter collective modes, 
 and multiple phases of the condensate. %,  in the condensed state \cite{Sigrist}.
 This is best illustrated by 
%While the spin-triplet pairing has not been well established in superconductors, the situation is quite clear in 
superfluid $^3$He, where  chargeless $^3$He atoms form spin-triplet  pairs (Ref. \cite{Legette,Tilley}). 
One of the important properties associated with  spin-triplet condensates is  spin-rotational symmetry breaking.
In  superfluid $^3$He, however, since there is no interaction to dictate the paired spins to point to a particular direction, the system preserves spin rotation symmetry. In superconductors such as UPt$_3$ (Ref.\cite{UPtReview}) or Sr$_2$RuO$_4$ (Refs. \cite{MaenoReview}), % or (TMTSF)$_2$PF$_6$ (Ref. \cite{Brown}),
it has been believed that the  Cooper pairs are in the spin-triplet state. Therefore,  evidence 
 for possible spin-rotation symmetry breaking had been  actively searched for, %particularly in UPt$_3$ and Sr$_2$RuO$_4$,
 since in  solids the spin pairs can favor a particular direction of the underlying lattice. Unfortunately, all such efforts have  not been very successful \cite{Tou,IshidaPRL}, which has been  interpreted as due to a weak spin-orbit coupling acting on the Cooper pairs \cite{MaenoReview2}. %}. %, the primary interaction to pin the spin direction of the Cooper pairs  
 This greatly hindered a firm establishment of the physics of spin-triplet superconductivity in these compounds \cite{UPtReview,Machida}.

Topological insulators (TIs) are a new class of quantum materials with strong spin-orbit coupling (SOC) and are characterized by topological invariants of the valence band \cite{Kane,ZhangSC,Ando}. 
%They feature  gapless  states in the surface while there is a  band gap  in the bulk .
 Bi$_2$Se$_3$ is one of the best studied  TIs \cite{Fang,Xia}, and  %where band-inversion occurs due to a strong spin-orbit coupling 
 most interestingly, Cu intercalation between the quintuple layers (see Fig. 1) %of Bi$_2$Se$_3$  
induces superconductivity below a maximum %transition temperature 
$T_c$=3.8 K (Ref. \cite{Hor}).
%The bulk nature of the superconductivity was assured by 
%A sizable jump in the specific heat at $T_c$ has been reported \cite{Kriener},  which points to the bulk nature of the superconductivity.
It was proposed that  %can be a topological superconductor with 
an  odd-parity, pseudo-spin triplet state %due to inter-orbit pairing between different %the two outer Se layers 
can be realized there \cite{Fu,Wan,Brydon}. 
Therefore, identifying the pairing state in Cu$_x$Bi$_2$Se$_3$  can not only enrich the physics of  spin-triplet superconductivity, but  also will lead to an establishment of a new frontier, % of superconductivity,
 topological superconductivity, % \cite{Fu,ZhangSC}, % the orbital part of the wave function of spin-triplet superconductivity in a centrosymmetric material necessarily has  
since odd parity has a % required by symmetry, which usually is topologically nontrivial 
%which, by analogy to topological insulators, has a pairing wave function of 
nontrivial topology  \cite{Fu,ZhangSC}.  
%The later task can have  additional practical importance since   the spin-polarized gapless surface  state of a topological superconductor  has potential applications in topological quantum computing \cite{Kitaev}. 

However, the experimental results on Cu$_x$Bi$_2$Se$_3$ from surface-sensitive probes    have been rather controversial \cite{Sasaki,STM,deViser}. Point contact spectroscopy  found a zero-bias conductance peak, which hints at an unconventional superconducting state \cite{Sasaki}. On the other hand, Scanning Tunneling Microscope measurement has found a U-shaped conductance spectrum which suggests a conventional gap \cite{STM}. Although it has been proposed that such seemingly contradicting results may stem from different doping level of the samples  that give rise to different shape of Fermi surface, and might be reconciled by considering the Dirac fermion states on the {\it surface} \cite{Mizushima}, bulk-sensitive and microscopic measurements are desired to resolve the issues. % Thus, the pairing symmetry in Cu$_x$Bi$_2$Se$_3$ is unsettled and measurements using bulk-sensitive probes are desired.

%By analogy, a superconductor whose 

%superconductors with the Fermi surface  enclosing an odd number of time reversal invariant  momenta and an odd-parity  gap  has a non-trivial topological invariant for the Cooper pairs wave function \cite{Fu,Sato},  and are called topological superconductors  whose surface state is necessarily gapless \cite{Fu,ZhangSC}. 
% whose surface harbors  zero-energy  Andreev bound states
%(Majorana fermions) \cite{ZhangSC} that are non Abelian \cite{Reed} and can be utilized  for quantum computing \cite{Kitaev}. 
%a superconducting material with non-trivial topological invariant for the Cooper pairs wave function can be topological, and
%For materials having a single Fermi surface enclosing the $\Gamma$($k$=0) point in the Brillouin zone, it has been shown that  an odd parity for the Cooper pairs wave function will make the materials topological \cite{Fu,Sato}. 
%

%Moreover, strong spin-orbit coupling will lift the spin degeneracy  of the gapless quasiparticle states on the surface. Such  novel  surface states are called helical Majorana fermions \cite{ZhangSC,Fu}, which 

%
%; such Majorana fermions are delocalized and dispersive, but when they re trapped in defects or in vortices, non-Abelian Majorana zero-mode may appear \cite{Reed, Fu, Hosur} and allow topological quantum computing \cite{Kitaev}.

%

\begin{figure}
\includegraphics[width=10cm]{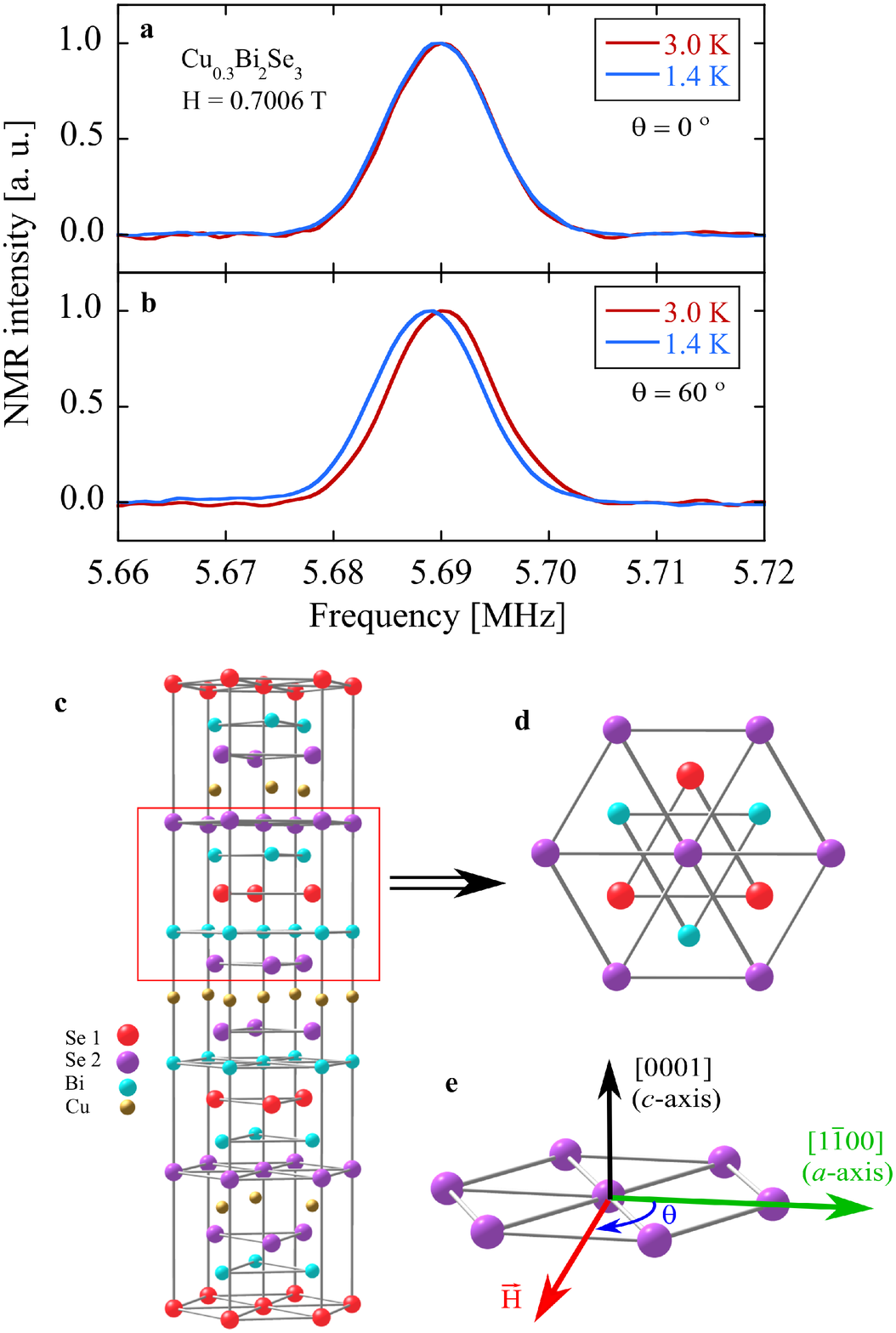}
\vspace{2cm}
\caption{\textbf{The NMR spectra of Cu$_{0.3}$Bi$_2$Se$_3$ with the magnetic field applied in the hexagonal plane}. 
{\textbf {a-b},} the $^{75}$Se NMR spectra at $T$=3.0 K (above $T_c$($H$)) and $T$=1.4 K (below $T_c$($H$)) for two representative field configurations.
 {\textbf c,} The crystal structure of Cu$_x$Bi$_2$Se$_3$. {\textbf d,} The top view (bird-eyes view) of the quintuple layer. {\textbf e,} illustration depicting the hexagonal plane
of the crystal and the magnetic-field angle  $\theta$. }
\label{spec}
\end{figure}

We performed $^{77}$Se nuclear magnetic resonance (NMR) measurements on Cu$_{0.3}$Bi$_2$Se$_3$ ($T_c$ = 3.4 K) to identify the pairing symmetry in the bulk of Cu$_x$Bi$_2$Se$_3$, since the Knight
shift in the superconducting state, which is proportional to the spin susceptibility through hyperfine coupling, is a good probe for the spin orientation of the Cooper pairs. 
%In the experiments we describe below, the skin depth that the electromagnetic wave penetrates is on the order of sub-millimeter \cite{Supple}. 
%by applying magnetic fields in various directions in the hexagonal basal plane as well as along the $c$-axis. 
Figure \ref{spec} shows typical $^{77}$Se NMR spectra for $H \parallel a$-axis in the hexagonal notation ([1$\bar{1}$00] direction in the rhombohedral notation) and $H \parallel \theta$ = 60$^{\circ}$ (rhombohedral [0$\bar{1}$10] direction), where $\theta$ is the angle measured from one of the $a$-axes in the hexagonal plane; our sample was rectangularly cut to have one of three  equivalent $a$-axes along the longer edge, which we define to be $\theta$ = 0$^{\circ}$.
Although there are two Se sites in the unit cell, the NMR transition line from them are not resolved, 
%This is also true even in the undoped Bi$_2$Se$_3$ \cite{Supple},  
due to small hyperfine coupling between the nuclear spin and electrons, and due to the fact that  $^{77}$Se with nuclear spin of 1/2 has no nuclear quadrupole moment which is commonly used to probe  different environment through couping to the electric field gradient.
The full width at half maximum %(FWHM) 
of the spectrum is about 10 kHz, which is much smaller than the bandwidth of the excitation RF field of $\sim$80 kHz \cite{Supple}.
Focusing on the change across $T_c$, %the superconducting transition,
one can see in Fig. \ref{spec} a clear shift of the spectrum below $T_c$ for $H \parallel \theta$ = 60$^{\circ}$, while the spectrum  changes only a little below $T_c$ for $H \parallel a$-axis.
It is emphasized that the spectrum shift of  $H \parallel \theta$ = 60$^{\circ}$ is uniform, which assures that there are no non-superconducting fraction that contributes to the observed spectrum \cite{Supple}.%, even though the superconducting shielding is not perfect  
%A similar situation was also seen in the superconductor Na$_x$CoO$_2$$\cdot$1.3H$_2$O \cite{Takada}.  There, even though the shielding fraction for a single crystal  was  less than 50\% \cite{LinCT}, %which also raised concern about the homogeneity of the sample , but 
%NMR measurements confirmed the good homogeneity and the bulk superconductivity \cite{zheng_PRB}. 

%We perform the $^{77}$Se nuclear magnetic resonance (NMR) measurements in Cu$_{0.25}$Bi$_2$Se$_3$ ($T_c$=3.8 K) with the magnetic field rotating in the hexagonal plane, and along the $c$-axis, respectively. 
%Figure \ref{spec}  shows the typical $^{77}$Se NMR spectra for $H \parallel a$-axis ([11$\bar{2}$0] direction), and $H \parallel \theta$=60$^{\circ} ([2\bar{1}\bar{1}0]$ direction), %and $H \parallel c$-axis, 
%respectively, where  $\theta$ is the angle from the $a$-axis in the hexagonal plane determined by Laue diffraction. A clear shift of the spectrum below $T_c$ can be seen for $H \parallel \theta$=60$^{\circ}$.
%Figure \ref{KvsT} shows a typical data set of the Knight shift as a function of temperature for the magnetic field applied along the $a$-axis and with  $\theta$=60$^{\circ}$, %$\deg$ 
%and along the $c$-axis, respectively. 

\begin{figure}
\includegraphics[width=12cm]{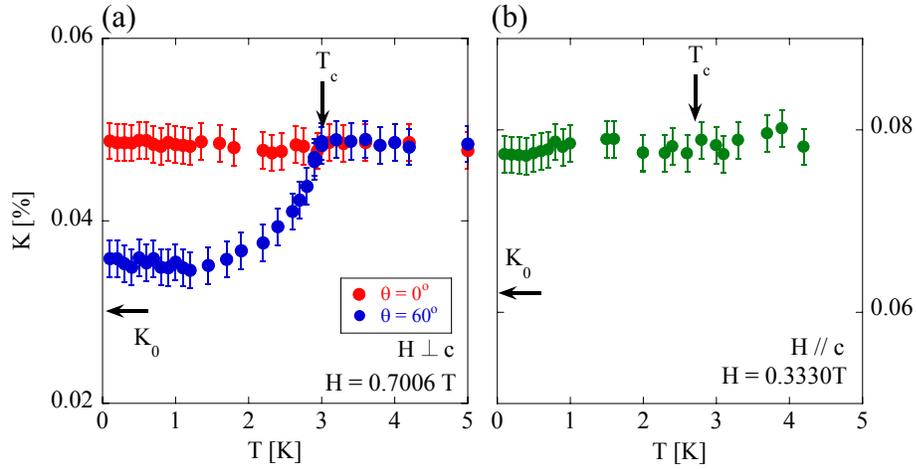}
\vspace{2cm}
\caption{ \textbf{Temperature dependence of the Knight shift with the magnetic field in the plane and along the $c$-axis.} \textbf{a}, the temperature dependence of the Knight shift in the hexagonal plane with $\theta$=0$^{\circ}$ and 60$^{\circ}$. 
%The fitting of the data for $H \parallel \theta$ = 60$^{\circ}$ assuming a mean-field temperature dependence reveals a gap $2\Delta(0\,{\rm K})$ = 3.54 $k_B T_c$, where $k_B$ is the Boltzmann constant. 
\textbf{b}, the Knight shift along the $c$-axis direction. %The sold curve and straight line are  guides to the eyes.  
In both cases, the horizontal arrow indicates the position of the shift for Bi$_2$Se$_3$.
Also, a correction of the diamagnetic field due to the shielding current around a vortex has been performed to account for the magnetic-field distribution in a sample for small applied fields (see Supplementary Information). %\cite{ZhengPRL2002}.  %The shift due to diamagnetism for $H\perpen$c-axis is small but is sizable for $H\parallel$c.  
 The Knight shift is calculated from the spectrum peak position which was obtained by fitting the data to a Gaussian function. The error bar represents the statistical error obtained from multiple spectrum measurements   at $T$=4.2 K. At low temperatures, the signal-to-noise ratio become better and the real error is smaller than the error bar shown.}
\label{KvsT}
\end{figure}

Figure \ref{KvsT} shows a typical data set of the Knight shift $K$ tensor as a function of $T$ %temperature 
 for $H$ along the $a$-axis and with $\theta$ = 60$^{\circ}$, as well as along the $c$-axis. 
%The $K$ consists of the part due to spin  susceptibility, $K_{\rm s}$, and the part  other than spin  susceptibility, which we denote as $K_{\rm orb}$, %which is $T$-independent, and the part due to  spin,  %through the hyperfine coupling,% between the electron  and the nuclear spins, 
%\begin{eqnarray}
%{\bm K}={\bm K}_{\rm s} + {\bm K}_{\rm orb}
%\\
%{\bm K}_{\rm s}={\bm A}_{\rm hf} {\bm \chi}_{\rm s}, 
%\end{eqnarray}
%where ${\bm A}_{\rm hf}$ is the hyperfine coupling constant and ${\bm \chi}_{\rm s}$ is the spin susceptibility. Thus, the Knight shift in the superconducting state is a good probe for the spin orientation of the Cooper pairs. 
%In conventional sense, $K_{orb}$ is due to electronic orbital susceptibility. In this work, $K_{orb}$ was determined from the NMR spectra of insulating Bi$_2$Se$_3$ \cite{Supple}.
The horizontal arrows in Fig. \ref{KvsT} indicate the position of the shift for insulating Bi$_2$Se$_3$ \cite{Supple}, which we denote as $K_{\rm 0}$ for convenience.  %position of the 
%orbital contribution to the Knight shift due to Van Vleck susceptibility, 
 % (See  Supplemental Materials ).
As seen in Fig. 2, 
a reduction of $K$ ($H\perp c$) is seen below $T_c$($H$) with a strong in-plane anisotropy
 (for more data, see \cite{Supple}). %See  Supplemental Materials for the data along all high-symmetry  directions  in the basal plane ). 

\begin{figure}
\includegraphics[width=12cm]{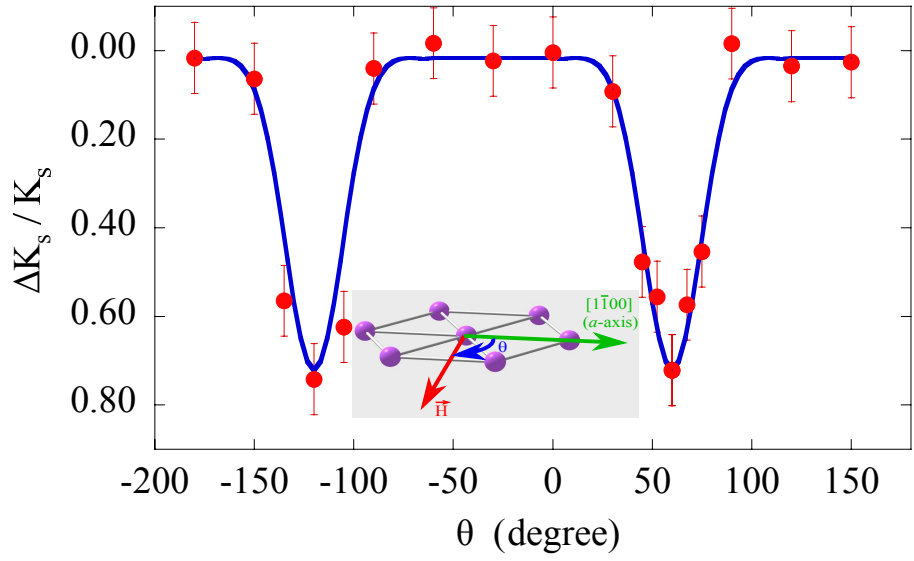}
\vspace{4cm}
\caption{\textbf{ Angle dependence of  the Knight shift reduction below $T_c$.} 
$\Delta K_s = K(3\,{\rm K})-K(1.4\,{\rm K})$, normalized by  $K_s = K(3\,{\rm K})-K_{\rm 0}$, is shown as a function of  $\theta$, the  angle between the in-plane magnetic-field and the $a$-axis.}
\label{Kvstheta}
\end{figure}

The key result of this work is shown in Fig. \ref{Kvstheta} 
which displays  the in-plane magnetic-field angle dependence of  the Knight shift reduction. The vertical axis is %the reduction of the in-plane spin Knight shift in the superconducting state, defined by %Knight shift ($\Delta K$) divided by its normal-state value ($K_N$), where 
$\Delta K_s / K_s$, where $\Delta K_s$= $K$($T$=3 K) - $K$ ($T$=1.4 K), and $K_s$ = $K$($T$=3 K)-$K_{\rm 0}$ which is due to (pseudo)spin associated with Cu-intercalation. %is the shift at $T$=3 K which is above $T_c$ in the applied field. 
Two sharp dips separated by 180$^{\circ}$ are observed, and they are located at $\theta$=60$^{\circ}$  and $\theta$=-120$^{\circ}$, which obviously breaks the three-fold rotation symmetry of the crystal. %Bi$_2$Se$_3$.
It should be  emphasized that, in the normal state above $T_c$, $K_s^c \sim K_s^a$ and the  shift is completely $\theta$-independent. % (see Fig. \ref{KvsT} and Supplemental Materials). 
Also, the doped Cu is randomly distributed as evidenced by the symmetric $^{63}$Cu-NMR lineshape \cite{Supple}, which indicates that Cu cannot  play a role in causing the anisotropy below $T_{\rm c}$.
Therefore, we conclude  that the spin rotation symmetry is broken in the hexagonal plane in the superconducting state of Cu$_{0.3}$Bi$_2$Se$_3$.
This is the first clear observation  for such spontaneous symmetry breaking in a superconductor.
%On the other hand, the Knight shift along the $c$-axis below $T_c$ is temperature independent  after correction of diamagnetism due to the field distribution in the vortex state (see Supplemental Materials).

The  broken spin rotation symmetry in the hexagonal plane and the invariance of $K_c$ below $T_c$ %and the $T$-independent $K_s^c$ 
%below $T_{\rm c}$ 
indicate that the Cooper pairs are in the pseudo-spin triplet state  in  Cu$_x$Bi$_2$Se$_3$. For a spin-triplet state, the gap 
%can have multiple components ($\Delta_{\uparrow \uparrow}$,    $\Delta_{\uparrow \downarrow}$, $\Delta_{\downarrow \uparrow}$ and $\Delta_{\downarrow \downarrow}$) and
 is  customarily described in terms of a three-component complex vector $\vec{d}(\vec{k})$=(
$d_x$($\vec{k}$), $d_y$($\vec{k}$), $d_z$($\vec{k}$)) to include the spin part of the wave function \cite{Balian,MaenoReview}. Usually, $\vec{d}(\vec{k})$ is defined as follows so that it will transform as a vector in spin space.
\begin{eqnarray}
\Delta(\vec{k})
=
\left(
\begin{array}{cc}
  \Delta_{\uparrow \uparrow}  & \Delta_{\uparrow \downarrow} \\
  \Delta_{\downarrow \uparrow} & \Delta_{\downarrow \downarrow} \\ 
 \end{array}%
\right)
=
\left(
\begin{array}{cc}
  -d_x+id_y & d_z \\
  d_z & d_x+id_y \\ 
 \end{array}%
\right)
\end{eqnarray} 
 
For the  case where $\vec{d}(\vec{k})\times \vec{d}^*(\vec{k})$=0 (which is called the unitary state) as is realized in the A or B phase of superfluid $^{3}$He \cite{Legette,Tilley}, the magnitude of the $\vec{d}(\vec{k})$-vector is simply the size of the superconducting gap, and the direction of the $\vec{d}(\vec{k})$-vector is perpendicular to the spin direction 
of the Cooper pairs. %In a usual  %simple 
%picture for intra-band spin-triplet  pairing,  %when $H$ is applied along the direction of the 
For $\vec{H}\parallel \vec{d}(\vec{k})$-vector, the spin susceptibility %(and the Knight shift) 
will decrease below $T_c$ as in a spin-singlet superconductor, while for %when $H$ is applied %along the spin quantization axis (which is perpendicular to the $\vec{d}(\vec{k})$-vector),
%perpendicular to the 
$\vec{H}\perp \vec{d}(\vec{k})$-vector,
 the spin susceptibility will stay unchanged across $T_c$. 
 Thus, in terms of the standard terminology, %for spin-triplet paring, 
the $\vec{d}(\vec{k})$-vector in Cu$_x$Bi$_2$Se$_3$ is  pinned to a particular direction in the crystal (perpendicular to one of the mirror planes).
%For inter-band pairing, the anisotropy due to the $\vec{d}(\vec{k})$-vector is the same as intra-band pairing, although additional features may appear  (see below) \cite{Fu,Hashimoto}.
%Therefore, in terms of  such terminology, the $\vec{d}(\vec{k})$-vector in Cu$_x$Bi$_2$Se$_3$ is pinned to one particular  crystal axis direction.
The  mechanism of the $\vec{d}(\vec{k})$-vector pinning to a particular direction ($\theta$ = 60$^{\circ}$) in the  three-fold-symmetric plane is unclear at the moment and
%probably due to some defects or alignments of the intercalated Cu atoms, whose details
 merits future investigation. In passing, 
we note that an  axis preference in the hexagonal plane was also  reported in the critical current  of UPt$_3$ below $T_c$ (Ref. \cite{Harlingen}). Also, the $^{195}$Pt Knight shift shows a tiny reduction (less than 1\% of the total shift) below $T_c$ along the $b$-axis ([1$\bar{1}$00])  and also along the $c$-axis, but no change was found along the $a$-axis ([11$\bar{2}$0]) \cite{Tou}.

%Looking into detail, we find that our result is a bit more complex than the simple behavior expected from intra-band spin-triplet  pairing, since the Knight shift along the $c$-axis and $a$-axis shows a tiny reduction below $T_{\rm c}$. 
%We here propose two possible explanations. % this tiny reduction. 
%Firstly, 
The  inter-orbital pairing theory proposed earlier for Cu$_x$Bi$_2$Se$_3$ is consistent with %may explain 
our result \cite{Fu}. Among the  proposed gap functions for which the spin susceptibility  has been calculated \cite{Hashimoto,Zocher},  %, as we discuss below.
%There are three Se layers in the QL unit (See Fig. 1), among them the two outer Se layers (Se2 sites) are centrosymmetric with respect to the inner-most Se layer (Se1 site). All of the Se layers possess a three-fold rotation symmetry and mirror  symmetry.  proposed that inter-orbital pairing between the two outer Se layers can give rise to a spin-triplet state. 
 the  pairing with electron $\vec{k}\uparrow$ ( $\vec{k}\downarrow$) on orbital 1 and electron $-\vec{k}\uparrow$  ($-\vec{k}\downarrow$) on orbital 2 %(or on orbital 1 and  on orbital 2)
 of the two Se2 layers 
%($c_{1k\uparrow}c_{2-k\uparrow},  c_{1k\downarrow}c_{2-k\downarrow}$), where $c_{1k\sigma}$ and $c_{2-k\sigma}$ are the annihilation operators for the orbital 1 and 2 of the two Se2 layers, 
has the $\vec{d}(\vec{k})$-vector with a main component  in the hexagonal plane \cite{Fu,Hashimoto}.
%and an additional small component along the $c$-direction , and %, and thus is consistent with our result. 
%the calculated spin susceptibility %of such a state 
%is consistent with our result \cite{Hashimoto}. Furthermore, 
According to the pseudo-spin description where the SOC was incorporated in the Hamiltonian which was used in the Bogoliubov-de Gennes equation to calculate the spin susceptibility in the superconducting state \cite{Hashimoto}, the spin susceptibility will only decrease along {\it one} particular direction in the hexagonal plane.  In such a direction, whether the spin susceptibility along the $\vec{d}(\vec{k})$-vector completely goes to zero at the $T$=0 K limit depends on the strength of the SOC \cite{Hashimoto}.
%
%in the band basis, a small amount of spin singlet component can be mixed due to SOC \cite{Hashimoto}, and this can explain the small, uniform reduction of the Knight shift along the directions away from $\theta$ = 60$^{\circ}$.  %, which lies in-between two of the $a$-axis directions. 
%As the second possibility,   the tiny reduction may  simply be a result of an underestimate of the shift due to diamagnetism in the vortex state \cite{Supple}. In the vortex state, the magnetic field inside a superconductor is inhomogeneous and a negative shift can occur at small applied field \cite{Supple}. %\cite{Zheng-PRL}. 
%The accuracy of the estimate of the diamagnetic shift relies on the parameters of penetration depth and coherence length.
%
%
%Another detailed aspect of the data deserves comment. 
In this regards, it is interesting to point out that even for $\theta$=60$^{\circ}$ or -120$^{\circ}$, the disappearance of $K_{\rm s}$ is not perfect but only about 80\%. This may be a signature of strong SOC of the system \cite{Hashimoto,Zocher}. 
%Qualitatively, such %intriguing
%  "residual" susceptibility in the $T$=0 K limit %of the superconducting state 
%is expected for a system with large SOC. 
Intuitively, 
this can be understood as the SOC contributing to spin-reversal scattering which makes spin not a good quantum number. This is also the reason why we use the term "pseudo-spin"  throughout this paper.
%Furthermore, according to  the pseudospin description for a band basis,  a small amount of "spin singlet" state  
%can be mixed due to spin-orbit coupling \cite{Hashimoto}, which  explains the small, uniform reduction of the Knight shift along the directions away from the $\theta$=60$^{\circ}$ direction. Finally,  the direction of $\theta$=60$^{\circ}$ is crystallographically equivalent to the $a$-axis determined by Laue diffraction, therefore the fact that the $\vec{d}(\vec{k})$-vector is pined in a particular direction in the otherwise isotropic hexagonal plane is probably due to some  defects whose details merit future investigation. 

Since inversion symmetry is preserved in Cu$_x$Bi$_2$Se$_3$, the pseudo-spin triplet state implies an odd parity. %for the spatial wave function 
%of the Cooper pairs %, irrespective of detail of theories 
%in the case of intra-orbit pairing. 
%It remains true 
%even in the case of inter-orbital pairing between two  outer Se layers in this compound \cite{Fu}. Here, %In the model of Fu and Berg, 
%although both the $E_u$ (odd parity) and $E_g$ representations are allowed in principle, %the  $E_g$ state can only have a weight of the order $\Delta/E_F$ which is negligibly small (here $E_F$ is the Fermi energy) \cite{Fu-PRB}. 
%Finally, we note that the Fermi surface of Cu$_x$Bi$_2$Se$_3$ encloses  only one time reversal invariant momentum at $\vec{k}$=0. 
%Furthermore, 
%the $E_g$ pairing has a $x^2-y^2$ symmetry and generates nodes along two axes ($x=\pm y$), which is inconsistent with the observed uni-axial spin anisotropy and suggests that the $E_u$ symmetry is most likely.
%It has been shown that for materials having a single Fermi surface enclosing the $\Gamma$ ($\vec{k}$ = 0) point in the Brillouin zone,  odd-parity superconducting state with a full gap will become topological \cite{Fu}.  %, irrespective of whether the superconducting gap is fully opened \cite{Fu} or has nodes (zeros) 
%Furthermore, it has  been proposed that  odd-parity superconductivity  with nodes can also be topological \cite{Sato}.  
Therefore, 
%although the gap structure  remains to be clarified in the future,  our finding shows that Cu$_x$Bi$_2$Se$_3$ %, whose Fermi surface  encloses  only one time-reversal-invariant momentum at $\vec{k}$=0, 
%fulfills the most severe conditions  for topological superconductivity. 
our results not only provide the first clear evidence  for spin-rotation symmetry breaking in a superconductor, but have a potential impact on %also could serve as a foundation toward 
establishing topological superconductivity which has an exciting prospect.
%can lead to an exiting prospect of   studying the  rich physics of a completely new class of superconductors. 

%\vspace{1cm}

%$^{+}$ Present address: RIKEN Center for Emergent Matter Science, 
%(CEMS)Strong Correlation Physics Division
%Strong Correlation Materials Research Team
%Hirosawa 2-1, Wako-shi, 
%Saitama 351-0198, Japan.

%$^{++}$ Present address: Department of Physics, Kyoto Sangyo University, Kyoto 603-8555, Japan.

%$^{+++}$ Present address: 

\textbf{Acknowledgments} We thank  A. Yamakage, M. Sato, L. Fu,  Y. Tanaka, K. Mizushima, Y. Yanase, J.P. Hu,  T. Xiang, Y. Maeno, K. Miyake and F. Yang for helpful discussion, and Y.S. Hor,  F. Iwase and K. Ueshima for participation in the initial stage of this work. This work was supported in part by MEXT research grants  (Innovative area ''Topological Quantum Phenomen'', No. 22103004 and ``Topological Materials Science'', No. 15H05852), JSPS grants(Nos. 24540320, 25220708 and 25800197), AFOSR (AOARD 124038) and by CAS.  

{\bf Author contributions}
 M.K and K.S synthesized and characterized the electrochemically-intercalated single crystals under the supervision of Y.A.
%. Y.S. Hor grew the melt-grown single crystals.  
K.M. and  G.Q.Z. performed  NMR measurements.  G.Q.Z integrated different pieces of the work
 and wrote the manuscript with inputs from all all co-authors. All authors discussed the results and interpretation.  %All authors have discussed the results and the interpretation.

{\bf Additional information}
Supplementary information is available in the online version of the paper.
Correspondence and requests for materials should be addressed to G.Q.Z.

{\bf Competing financial interests}
The authors declare no competing financial interests.

\clearpage

\clearpage

\clearpage

\end{document}